\documentclass[11pt]{article}
\usepackage{amsmath,amsfonts,slashed}
\usepackage{epsfig,epsf}
\usepackage{hyperref}
\usepackage{float}
\usepackage[dvips]{color}
\usepackage[normalem]{ulem}
\usepackage{graphicx}
\usepackage[tight,footnotesize]{subfigure}
\usepackage{hyperref}
\usepackage{slashed}
\def \met  {\mbox{${E\!\!\!\!/_T}$}}
\begin{document}

\renewcommand*{\thefootnote}{\fnsymbol{footnote}}
\begin{center}
{\Large \bf Exploring non minimal Universal Extra Dimensional Model at the LHC }\\
\vspace*{1cm} 
{\sf Nabanita Ganguly\footnote{email: nabanita.rimpi@gmail.com}},
{{\sf Anindya Datta\footnote{email: adphys@caluniv.ac.in}}  
} \\  
\vspace{10pt}  
{\small {\em Department of Physics, University of Calcutta,  
92 Acharya Prafulla Chandra Road,\\
Kolkata 700009, India}}
 
\normalsize  
\end{center}

\vspace{5mm}
\begin{abstract}\vspace*{10pt}

We study the collider phenomenology of non minimal universal extra dimensional (nmUED) model in the context of the Large Hadron Collider at CERN.  nmUED is an incarnation of the Standard Model in $(4+1)$  space-time dimensions compactified on an $\bf{S_1}/\bf{Z_2}$ orbifold supplemented with  boundary localized operators with unknown coefficients. These coefficients parametrize the radiative corrections which are somehow arbitrary due to the lack of knowledge of the cut-off scale of such effective theory.  It is possible to tune the masses and couplings of the Kaluza Klein (KK) excitations by tuning these parameters. Two scenarios with different mass hierarchies among the  KK-excitations have been  considered. A detailed study of production cross-sections and different decay modes of KK-particles are also presented. We calculate the correlated bounds on masses of KK-particles using the LHC $nl + mj + \met$ data (in case of production of strongly interacting particles) and $3l + \met$ data (in case of electroweak productions). We work in a KK-parity conserving scenario where the lightest KK-particle is a potential dark (DM) matter candidate. We use the additional constraints coming from the observed DM relic density of the universe to identify the allowed parameter space. The current status of a nmUED model in the light of DM direct detection data is also examined. The present investigation reveals that production and subsequent leptonic decays of KK-electroweak gauge bosons as massive as 1 TeV, lead to observable trilepton signature with a luminosity of 1 ab$^{-1}$ which will be realized in near future.

\vskip 5pt \noindent  
\texttt{PACS Nos:~11.10.Kk, 14.80.Rt, 95.35.+d, 13.85.-t
  } \\  
\texttt{Key Words:~~Universal Extra Dimension, Dark matter, Kaluza-Klein, LHC} 
\end{abstract}
\renewcommand{\thesection}{\Roman{section}}  
\setcounter{footnote}{0}  
\renewcommand{\thefootnote}{\arabic{footnote}}
\section{Introduction}
Discovery of the Higgs boson at the LHC \cite{Aad:2012tfa, Chatrchyan:2012xdj} has  once more undoubtedly proved the triumph  of the Standard model (SM). However, a framework beyond the SM (BSM) is a need of time more than ever because of mainly two very pressing experimental evidences. One is of course, the signatures  of tiny but non zero neutrino mass \cite{Hirata:1988uy, Hirata:1989zj, Casper:1990ac, Fukuda:1998mi, Ahmad:2001an, Eguchi:2002dm, Aliu:2004sq, Hosaka:2005um, Michael:2006rx, Arpesella:2007xf} and the second is the experimental signals in support of Dark Matter (DM) \cite{Olive:2005qz, Baer:2008uu, Arrenberg:2013rzp}. Models of extra dimensions \cite{Antoniadis:1990rq,ArkaniHamed:1998rs,Antoniadis:1998ig,Randall:1999ee,Randall:1999vf}, which have been put forward as an alternative for supersymmetry to solve the hierarchy problem in the SM also  provide a natural framework for models of DM  and neutrino masses. In particular, we are interested in a variant of extra dimensional model, namely the Universal Extra Dimensional (UED) model which is nothing but an incarnation of the SM in $(4+1)$ space-time dimensions. In a minimal model, the extra space-like dimension is compactified on a circle of radius $R$, $R^{-1}$ being a energy scale above which the new dynamics beyond the SM would reveal itself. UED is hallmarked by the presence of Kaluza Klein (KK) modes of the familiar SM particles. The masses of the KK-modes at n$^{th}$ the KK-level are $n R^{-1}$. n is called the KK-number which is nothing but quantised momentum in $y$-direction. A $\bf{Z}_2$ ($y \rightarrow -y$) symmetry is imposed on the action to get chiral fermions and at the same time to remove some unwanted degrees of freedom at zero KK-level which is identified with SM. This method is known as orbifolding. Orbifolding leads to a KK-parity defined as $(-1)^n$ at the n$^{th}$ KK-level. The lightest one among all the KK-modes, which is massive and generally weakly interacting and stable due to  the KK-parity, could be a candidate for DM \cite{Servant:2002hb, Servant:2002aq, Belanger:2010yx, Kong:2005hn, Burnell:2005hm}. The masses of KK-excitations of SM particles at a given KK-level are closely spaced. Radiative correction to the masses are important and could be incorporated easily by following the refs. \cite{Georgi:2000ks, Cheng:2002iz}. However, in an effective theory the radiative corrections are sensitive to an (unknown) energy cut-off of the theory\footnote{It has been shown that one loop effects in a theory with one extra space-like dimension are only log sensitive to the cut-off \cite{Dey:2004gb}.} and thus somehow arbitrary in nature. As a consequence of orbifolding boundary conditions, momentum conservation along the 5$^{th}$ (extra) dimension is lost. This in turn forces the non-zero mass corrections to be localized at the boundaries of the extra dimension.  The minimal version of the UED model \cite{Cheng:2002iz, Cheng:2002ab} takes some special assumption resulting into vanishing boundary radiative corrections at the cut-off scale\footnote{The cut-off dependence of different physical observables is studied in \cite{Freitas:2018bag}.}.

A more general version of the model \cite{Dvali:2001gm, Carena:2002me, delAguila:2003bh, delAguila:2003kd, delAguila:2003gv, Schwinn:2004xa, Flacke:2008ne, Datta:2012xy} treats the coefficients of the boundary localized terms (BLTs) generated by radiative corrections as free parameters of the theory. This framework is known as non minimal Universal Extra Dimensional (nmUED) model in the sense that the coefficients of the boundary localized operators that may arise due to radiative corrections are assumed to be arbitrary free parameters. The allowed range of values of these parameters could be fixed from a unitarity analysis of this theory \cite{Jha:2016sre}. Furthermore, a lot of effort has been put in the investigation of such model from the perspective of DM \cite{Datta:2013nua, Dasgupta:2018nzt}, $Z$-boson decay to $b$-quarks \cite{Jha:2014faa}, rare decays of top quarks \cite{Dey:2016cve, Chiang:2018oyd}, gauge fixing \cite{Datta:2014sha}, decays of $B$-meson \cite{Datta:2015aka, Datta:2016flx} and more recently $R(D^{(\star)})$ anomalies \cite{Biswas:2017vhc}. In \cite{Flacke:2017xsv}, authors have done a detailed analysis of KK-dark matter in the light of various experimental constraints. Productions of KK-excitations of the strongly interacting KK-quarks and KK-gluons at the LHC have been investigated in \cite{Datta:2012tv}. Signatures of the minimal UED model (mUED) has been put forward in the context of LHC and constraints on mUED parameter space has also been derived from comparing the LHC data with signal strength in mUED model \cite{Choudhury:2016tff}. However no serious effort has so far put in to derive such bounds on nmUED set of parameters from the LHC data. In this paper we will investigate the prospects of  discovering signatures of nmUED model comprising of jets and leptons at the LHC. We also estimate the strength of trilepton signal in association with missing energy ($\met$) in the nmUED model and compare with the available LHC data to constrain nmUED parameters. 

The plan of the paper is following. In section \ref{models}, we will briefly discuss the model with emphasis on the choices of BLT coefficients leading to  different mass hierarchies among the KK-excitations in the first KK-level. Section \ref{prdctn} will be devoted to the productions and decays of strong and electroweak (EW) KK-excitations at the LHC running at 13 TeV. In section \ref{method}, we discuss the methodology adapted for our analysis and also present the results. Finally we conclude in section \ref{conclusion}.

\section{Description of model}
\label{models}
In this section we briefly describe the nmUED scenario. To start with, we will keep our discussion more general in nature. In Table \ref{tab1}, we show different BLT coefficients and the corresponding KK-masses that will appear in the following discussion. However, in our analysis several special choices of the input parameters will be assumed to investigate different mass hierarchies among the KK-excitations of $n=1$ level.  These finer details of choices of the parameters will be emphasized as we go along. A more detailed discussion on the  model and its technicalities are available in refs. \cite{Dvali:2001gm, Carena:2002me, delAguila:2003bh, delAguila:2003kd, delAguila:2003gv, Schwinn:2004xa, Flacke:2008ne, Datta:2012xy}.

\begin{table}[H]
\begin{center}
\begin{tabular}{|c|c|c|}
\hline
\hline
BLT coefficients &Symbol &KK-mass \\
\hline
$U(1)_Y$ &$r_B$ &$M_B^{(1)}$   \\
$SU(2)_L$ &$r_W$ &$M_W^{(1)}$  \\
$SU(3)_C$ &$r_{gl}$ &$M_{gl}^{(1)}$  \\
Quark &$r_q$ &$M_q^{(1)}$  \\
Lepton &$r_l$ &$M_l^{(1)}$   \\
Scalar &$r_{\phi}$ &$M_{\phi}^{(1)}$ \\
Quark Yukawa &$r_q^Y$ &- \\
Lepton Yukawa &$r_l^Y$ &-\\
\hline
\hline
\end{tabular}
\end{center}
\caption{Notations for different BLT coefficients and the corresponding KK-masses.}
\label{tab1}
\end{table} 

We start with the fermion sector of KK-spectrum. The 5 dimensional action for KK-quarks in the presence of BLTs is given by
\begin{eqnarray}
\label{lagquark}
\mathcal{S}_{quark} &=& \int d^4 x\;\int_{0}^{\pi R} dy \Big[\overline{Q} i\Gamma^{M} \mathcal{D}_{M} Q + r_q \{ \delta(y) + \delta(y-\pi R) \} \overline{Q} i\gamma^{\mu} \mathcal{D}_{\mu} P_L Q \nonumber \\ 
 & & + \overline{U} i\Gamma^{M} \mathcal{D}_{M} U + r_q \{ \delta(y) + \delta(y-\pi R) \} \overline{U} i\gamma^{\mu} \mathcal{D}_{\mu} P_R U \nonumber \\ 
 & & + \overline{D} i\Gamma^{M} \mathcal{D}_{M} D + r_q \{ \delta(y) + \delta(y-\pi R) \} \overline{D} i\gamma^{\mu} \mathcal{D}_{\mu} P_R D\Big] 
\end{eqnarray}
where $Q(x,y)$ is the $SU(2)_L$ doublet quark field and $U(x,y), D(x,y)$ are the up and down-type $SU(2)_L$ singlet quark fields respectively. $M,N = 0,1,2,3,4$ stand for 5 dimensional Lorentz indices with the metric convention $g_{MN} \equiv {\rm diag}(+,-,-,-,-)$. $D_M$ is the usual covariant derivative in 5 dimension defined as $D_M\equiv\partial_M-i\widetilde{g}_2 W_M^a T^a-i \widetilde{g}_3 G_M^a \lambda^a-i \widetilde{g}_1 B_M Y$ where $\widetilde{g_1}$, $\widetilde{g_2}$ and $\widetilde{g_3}$ are the 5 dimensional gauge couplings, $Y$, $T^a$ and $\lambda^a$ are the generators, $B_M$, $W_M^a$ and $G_M^a$ are the gauge fields for the gauge groups $U(1)_Y$, $SU(2)_L$ and $SU(3)_C$ respectively. $\Gamma_M$ is the 5 dimensional representation of the Clifford algebra given by $\Gamma_\mu = \gamma_\mu$; $\Gamma_4 = i \gamma_5$.

The 5 dimensional quark fields can be expanded in terms of 4 dimensional fields as follows
\begin{eqnarray}
\label{quarkexpnsn}
Q(x,y) &=&   \left( \begin{array}{c} Q_{u, L}^{(0)}(x) f_L^{(0)}(y) + \sum^{\infty}_{n=1} (Q_{u, L}^{(n)}(x) f_L^{(n)}(y) + Q_{u, R}^{(n)}(x) g_L^{(n)}(y)) \\ Q_{d, L}^{(0)}(x) f_L^{(0)}(y) +\sum^{\infty}_{n=1} (Q_{d, L}^{(n)}(x) f_L^{(n)}(y) + Q_{d, R}^{(n)}(x) g_L^{(n)}(y)) \end{array} \right) \nonumber \\
U(x,y) &=& u_{R}^{(0)}(x) g_R^{(0)}(y) + \sum^{\infty}_{n=1}  (u_{R}^{(n)}(x) g_R^{(n)}(y) + u_{L}^{(n)}(x) f_R^{(n)}(y)) \nonumber \\
D(x,y) &=& d_{R}^{(0)}(x) g_R^{(0)}(y) + \sum^{\infty}_{n=1}  (d_{R}^{(n)}(x) g_R^{(n)}(y) + d_{L}^{(n)}(x) f_R^{(n)}(y))
\end{eqnarray} 
where $Q_{u,L}(x)$ $(Q_{d,L}(x))$ is the up-(down-)type doublet and $u_{R}(x)$ $(d_{R}(x))$ is the up-(down-)type singlet 4 dimensional quark fields. A 5 dimensional Lagrangian involving leptons can be similarly written down keeping in mind that right handed neutrinos are not present 
in our model as we are not interested in explaining neutrino masses at this moment. 

If a fermion has a KK-mass $M_f^{(n)}$ and $r_{f}$ is the corresponding BLT coefficient, it can be shown that $M_f^{(n)}$ satisfies a transcendental equation of the following form
\begin{eqnarray}
 r_{f} M_f^{(n)}= \left\{ \begin{array}{rl}
         -2\tan \left(\frac{M_f^{(n)}\pi R}{2}\right) &\mbox{when $n$ is even,}\\
          2\cot \left(\frac{M_f^{(n)}\pi R}{2}\right) &\mbox{when $n$ is odd.}
          \end{array} \right.   
          \label{transcendental}      
 \end{eqnarray}
where $M_f^{(n)} = M_q^{(n)}, M_l^{(n)}$ and $r_{f} = r_{q}, r_{l}$ for KK-quarks and KK-leptons respectively.

The $y$-profile of the 5 dimensional fermion field is given by
\begin{eqnarray}
f_{L}^{(n)} = g_{R}^{(n)} = N_{fn} \left\{ \begin{array}{rl}
                \displaystyle \frac{\cos(M_f^{(n)} \left (y - \frac{\pi R}{2}\right))}{C_{f_{n}}}  &\mbox{for $n$ even,}\\
                \displaystyle \frac{{-}\sin(M_f^{(n)} \left (y - \frac{\pi R}{2}\right))}{S_{f_{n}}} &\mbox{for $n$ odd.}
                \end{array} \right.
                \label{flgr}
\end{eqnarray}
and
\begin{eqnarray}
g_{L}^{(n)} = f_{R}^{(n)} = N_{fn} \left\{ \begin{array}{rl}
                \displaystyle \frac{\sin(M_f^{(n)} \left (y - \frac{\pi R}{2}\right))}{C_{f_{n}}}  &\mbox{for $n$ even,}\\
                \displaystyle \frac{\cos(M_f^{(n)} \left (y - \frac{\pi R}{2}\right))}{S_{f_{n}}} &\mbox{for $n$ odd.}
                \end{array} \right.
                \label{glfr}
\end{eqnarray} 
with
\begin{equation}
C_{f_{n}} = \cos\left( \frac{M_f^{(n)} \pi R}{2} \right)\,\, , 
\,\,\,\,
S_{f_{n}} = \sin\left( \frac{M_f^{(n)} \pi R}{2} \right)
\end{equation}

The wave functions along the $y$-direction satisfy the orthonormality conditions
\begin{equation}
\int dy \left[1 + r_{f}\{ \delta(y) + \delta(y - \pi R)\}
\right] ~f_L^{(n)}(y) ~f_L^{(m)}(y) = \int dy ~g_L^{(n)}(y) ~g_L^{(m)}(y) = \delta^{n m}
\label{ortho}
\end{equation}

Substituting eqs. \ref{flgr}, \ref{glfr} into eq. \ref{ortho}, one finds the normalization constant as
\begin{equation}
N_{fn} = \sqrt{\frac{2}{\pi R}}\left[ \frac{1}{\sqrt{1 + \frac{r_f^2 M_f^{(n)2}}{4} 
+ \frac{r_f}{\pi R}}}\right]
\label{normalisation}
\end{equation}

Next we turn our attention to the gauge and scalar sectors of the theory. The general forms of 5 dimensional actions with BLTs in this case are given by
\begin{eqnarray}
\label{su2}
\mathcal{S}_{SU(2)} &=& \int d^4 x\; \int_{0}^{\pi R} dy\Big[-\frac{1}{4}{\cal W}^{MNa}{\cal W}_{MN}^{a} -\frac{r_{W}}{4}\{ \delta(y) + \delta(y - \pi R)\}{\cal W}^{\mu \nu a}{\cal W}_{\mu \nu}^{a} \Big] \nonumber\\
\label{u1}
\mathcal{S}_{U(1)} &=& \int d^4 x\; \int_{0}^{\pi R} dy\Big[-\frac{1}{4}{\cal B}^{MN}{\cal B}_{MN} -\frac{r_{B}}{4}\{ \delta(y) + \delta(y - \pi R)\} {\cal B}^{\mu \nu } {\cal B}_{\mu \nu} \Big] \nonumber\\
\label{su3}
\mathcal{S}_{SU(3)} &=& \int d^4 x\; \int_{0}^{\pi R} dy\Big[-\frac{1}{4}{\cal G}^{MNa}{\cal G}_{MN}^{a} -\frac{r_{gl}}{4}\{ \delta(y) + \delta(y - \pi R)\}{\cal G}^{\mu \nu a}{\cal G}_{\mu \nu}^{a} \Big] \nonumber\\
\label{higgs}
\mathcal{S}_{\Phi} &=& \int d^4 x\; \int_{0}^{\pi R} dy \Big[\left(D^{M}\Phi\right)^{\dagger}\left(D_{M}\Phi\right) + \widetilde{\mu}^2 \Phi^{\dagger} \Phi - \frac{\widetilde{\lambda}}{4} (\Phi^{\dagger} \Phi)^2 \nonumber \\
 & & +  r_{\phi}\{ \delta(y) + \delta(y - \pi R)\} (\left(D^{\mu}\Phi\right)^{\dagger}\left(D_{\mu}\Phi\right) + \widetilde{\mu}^2 \Phi^{\dagger} \Phi - \frac{\widetilde{\lambda}}{4} (\Phi^{\dagger} \Phi)^2) \Big]
\end{eqnarray}
where $\Phi(x,y) = \left(\begin{array}{c} \Phi^{+}(x,y) \\ \Phi^{0}(x,y) \end{array} \right)$ is the Higgs doublet, `a' is the gauge index with a = 1,2,3 for $SU(2)_L$ and a = 1,...,8 for $SU(3)_C$. $\widetilde{\mu}$, $\widetilde{\lambda}$ are the Higgs mass parameter and quartic coupling respectively in 5 dimension. The field strength tensors are given by
\begin{eqnarray}
\label{fieldstrenght}
{\cal W}_{MN}^a &\equiv& (\partial_M W_N^a - \partial_N W_M^a+\widetilde{g}_2 \epsilon^{abc}W_M^bW_N^c) \nonumber  \\
{\cal G}_{MN}^a &\equiv& (\partial_M G_N^a - \partial_N G_M^a+\widetilde{g}_3 f^{abc}G_M^b G_N^c) \nonumber  \\
{\cal B}_{MN} &\equiv& \partial_M B_N - \partial_N B_M
\end{eqnarray}

The $y$-dependent wave functions for 5 dimensional KK-scalar and KK-gauge boson fields are given by eqs. \ref{flgr} and \ref{glfr} with appropriate replacements for $M_f^{(n)}$ and $r_f$. 

The mass matrix involving $W_{\mu}^{3 (n)}$ and $B_{\mu}^{(n)}$ takes the following form 
\begin{equation}
\label{mixingmatrix}
\mathcal{M}_{W^3B}^{(n)} =
\begin{pmatrix}
\frac{g_2^2 v^2}{8} + M_{W}^{(n) 2} & -\frac{g_1 g_2 v^2}{8} \sqrt{\frac{\pi R + r_B}{\pi R + r_W}} I^{(n)} _{W^3B} \\ -\frac{g_1 g_2 v^2}{8} \sqrt{\frac{\pi R + r_B}{\pi R + r_W}} I ^{(n)} _{W^3B} & \frac{g_1^2 v^2}{8} \frac{\pi R + r_B}{\pi R + r_W} I ^{(n)} _{BB} + M_{B}^{(n) 2}
\end{pmatrix}
\end{equation}
where $M_{W}^{(n)}$ and $M_{B}^{(n)}$ are the KK-masses at the $n^{th}$ level corresponding to BLT coefficients $r_W$ and $r_B$ respectively. $I_{W^3B}^{(n)}$, $I_{BB}^{(n)}$ are overlap integrals given by 
\begin{eqnarray}
\label{integrals}
I ^{(n)}_{W^3B} &=& \int_0 ^{\pi R}
dy  \;[1+r_{W}\{\delta(y)+\delta(y-\pi R)\}]f_{L,W}^{(n)} f_{L,B}^{(n)} \nonumber \\
I ^{(n)} _{BB} &=& \int_0 ^{\pi R}
dy  \;[1+r_{B}\{\delta(y)+\delta(y-\pi R)\}]f_{L,B}^{2 (n)}
\end{eqnarray}

Upon diagonalization of the matrix given in \ref{mixingmatrix}, the gauge eigenstates $W_{\mu}^{3 (n)}, B_{\mu}^{(n)}$ are found to be related to the mass eigenstates $Z_{\mu}^{(n)}$ and $A_{\mu}^{(n)}$ through the following relations
\begin{eqnarray}
\label{rotations}
W_{\mu}^{3 (n)} = \cos\theta_W^{(n)} Z_{\mu}^{(n)} + \sin\theta_W^{(n)} A_{\mu}^{(n)} \nonumber \\
B_{\mu}^{(n)} = - \sin\theta_W^{(n)} Z_{\mu}^{(n)} + \cos\theta_W^{(n)} A_{\mu}^{(n)} 
\end{eqnarray}   
where the mixing angle $\theta_W^{(n)}$ corresponding to the $n^{th}$ level is given by
\begin{equation}
\label{mixingangle}
\tan\theta_W^{(n)} = \frac{\frac{g_1 g_2 v^2}{4} \sqrt{\frac{\pi R + r_B}{\pi R + r_W}} I ^{(n)} _{W^3B}}{\frac{g_2^2 v^2}{8} + M_{W}^{(n) 2} - M_{B}^{(n) 2} - \frac{g_1^2 v^2}{8}\frac{\pi R + r_B}{\pi R + r_W} I_{BB}^{(n)}}
\end{equation} 

The KK-masses of the physical states $Z_{\mu}^{(n)}$ and $A_{\mu}^{(n)}$ are given by the eigenvalues of the matrix ${M}_{W^3B}^{(n)}$. Also note that the mixing between $W_{5}^{\pm (n)}$ and $\Phi^{\pm (n)}$ gives rise to a tower of physical charged Higgs fields $H^{\pm (n)}$ while the mixing between $Z_5^{(n)}$\footnote{This state arises from the mixing between $W_5^{3 (n)}$ and $B_5^{(n)}$ which are the 5$^{th}$ components of KK-excitations of EW gauge bosons.} and $\chi^{(n)}$ gives rise to a tower of physical neutral pseudo scalar Higgs fields $A^{(n)}$\footnote{$\Phi^{\pm (n)}$ and $\chi^{(n)}$ are KK-excitations of charged and scalar components respectively of the Higgs doublet $\Phi(x,y)$.}. The orthogonal combinations to the physical charged scalars and pseudo scalars are Goldstone bosons which decouple from the theory with an unitary gauge choice to which we adhere in the following. The physical mass of a KK-particle at the $n^{th}$ KK-mode is given by $m^{(n)} = \sqrt{M_{SM}^2 + M_{KK}^{(n)2}}$ where $M_{SM}$ is the zero mode mass and $M_{KK}^{(n)}$ is the KK-mass at the $n^{th}$ level which is a solution of the transcendental equation of the form \ref{transcendental}.

Finally, we come to the Yukawa sector. The 5 dimensional Yukawa action for quarks in presence of BLTs is given by
\begin{eqnarray}
\label{quarkyukawa}
\mathcal{S}_{Y}^{quark} &=& - \int d^4 x \int_{0}^{\pi R} dy\Big[\widetilde{y_u} \bar{Q} \widetilde{\Phi} U + \widetilde{y_d} \bar{Q} \Phi D + r_{Y}^{q}\{ \delta(y) + \delta(y - \pi R)\} \nonumber \\
& & (\widetilde{y_u} \bar{Q_L} \widetilde{\Phi} U_R + \widetilde{y_d} \bar{Q_L} \Phi D_R) + \textrm{h.c.} \Big]  
\end{eqnarray}   
where $\widetilde{\Phi} \equiv i\tau^2 \Phi^\ast$\footnote{where $\tau^2$ is the usual Pauli matrix.}, $\widetilde{y_u}$ $(\widetilde{y_d})$ stands for up-(down-)type 5 dimensional Yukawa coupling and $r_Y^q$ is the BLT coefficient for the corresponding Yukawa interactions. Similarly for leptons, the corresponding action in (4+1) dimension would be
\begin{eqnarray}
\label{leptonyukawa}
\mathcal{S}_{Y}^{lepton} &=& - \int d^4 x \int_{0}^{\pi R} dy\Big[\widetilde{y_l} \bar{L} \Phi E + r_{Y}^{l}\{ \delta(y) + \delta(y - \pi R)\}\widetilde{y_l} \bar{L_L} \Phi E_R + \textrm{h.c.} \Big]  
\end{eqnarray} 
where $\widetilde{y_l}$ is the 5 dimensional lepton Yukawa coupling and $r_Y^l$ is the BLT coefficient of the corresponding interactions. For simplicity of calculations, we set the quark (lepton) Yukawa BLT coefficient equal to the quark (lepton) BLT coefficient i.e. $r_Y^q = r_q$ and $r_Y^l= r_l$\footnote{For the choice of unequal Yukawa and fermion BLT coefficients the corresponding fermion mixing matrix will have a complicated form and will lead to mixing among KK-fermions of different KK-levels. For details, see \cite{Jha:2014faa}.}. We will stick to this choice in all our calculations. 

We conclude the model description by stating the relations involving 5 dimensional parameters and their SM counterparts. They can easily be derived by substituting all the field expansions given in eqs. \ref{quarkexpnsn}, \ref{quarkyukawa}, \ref{leptonyukawa} into the corresponding Lagrangians and comparing the zero mode with the SM Lagrangian. We enlist them below.
\begin{eqnarray}
\label{modifiedrelations}
\widetilde{g_1} &=& g_1 \sqrt{\pi R + r_B} \nonumber \\
\widetilde{g_2} &=& g_2 \sqrt{\pi R + r_W} \nonumber \\
\widetilde{g_3} &=& g_3 \sqrt{\pi R + r_{gl}} \nonumber \\
\widetilde{y} &=& y \sqrt{\pi R + r_W} \nonumber \\
\end{eqnarray}
where $\widetilde{y}$ stands for general Yukawa coupling.

\begin{table}[H]
\begin{center}
\begin{tabular}{|c|c||c|}
\hline
\hline
Sector &Scenario A &Scenario B \\
\hline
Gauge &$r_B = r_W = r_{EW}, r_{gl}$ &$r_B \neq r_W, r_{gl}$  \\
\hline
Fermion &$r_q = r_l = r_f$ &$r_q \neq r_l$ \\
\hline
Scalar &$r_{\phi} = r_W$ &$r_{\phi} = r_W$ \\
\hline
Yukawa &$r_q^Y = r_l^Y = r_f$ &$r_q^Y = r_q, r_l^Y = r_l$ \\
\hline
Hierarchy &$ r_{gl} < r_q  (r_l) < r_B (= r_W)$  &  $ r_{gl} < r_q  < r_W <  r_l  <  r_B$ \\
\hline
\hline
\end{tabular}
\end{center}
\caption{Definitions of two different scenarios used in the analysis.}
\label{tab2}
\end{table} 

The above discussion about the nmUED model is more or less general in nature. In our analysis we consider two different scenarios to study signals at the LHC. The definition of each scenario is given in Table \ref{tab2}. Note that in case of scenario A, several expressions involving masses and mixing of $U(1)_Y$ and $SU(2)_L$ gauge bosons become simpler because of the equality of the involved BLT coefficients. As for example, the mass matrix in eq. \ref{rotations} reduces to the simple form of the corresponding mass matrix in the SM. The hierarchies among BLT coefficients in each type of model are also shown. Note that we always take $r_B$ to be the largest so that in a KK-parity conserving scenario $\gamma^{(1)}$ becomes the lightest KK-particle (LKP) and hence a DM candidate\footnote{The phenomenological impact of non conservation of KK-parity has been studied in \cite{Datta:2013lja, Shaw:2014gba, Shaw:2017whr}.}.

\section{Production and decay modes of KK particles}
\label{prdctn}

In this section we discuss the relevant tree level processes leading to productions of level 1 EW and strongly interacting KK-particles of nmUED at the LHC. One of our main objectives in this work is to see how much parameter space of nmUED has already been explored by the presently available LHC data. For this, a knowledge of production cross-sections of the KK-particles is very important. We also discuss various decay modes of these KK-particles that depend crucially on the hierarchy among BLT coefficients.

\begin{figure}[H]
\centering
\includegraphics[width=0.6\textwidth]{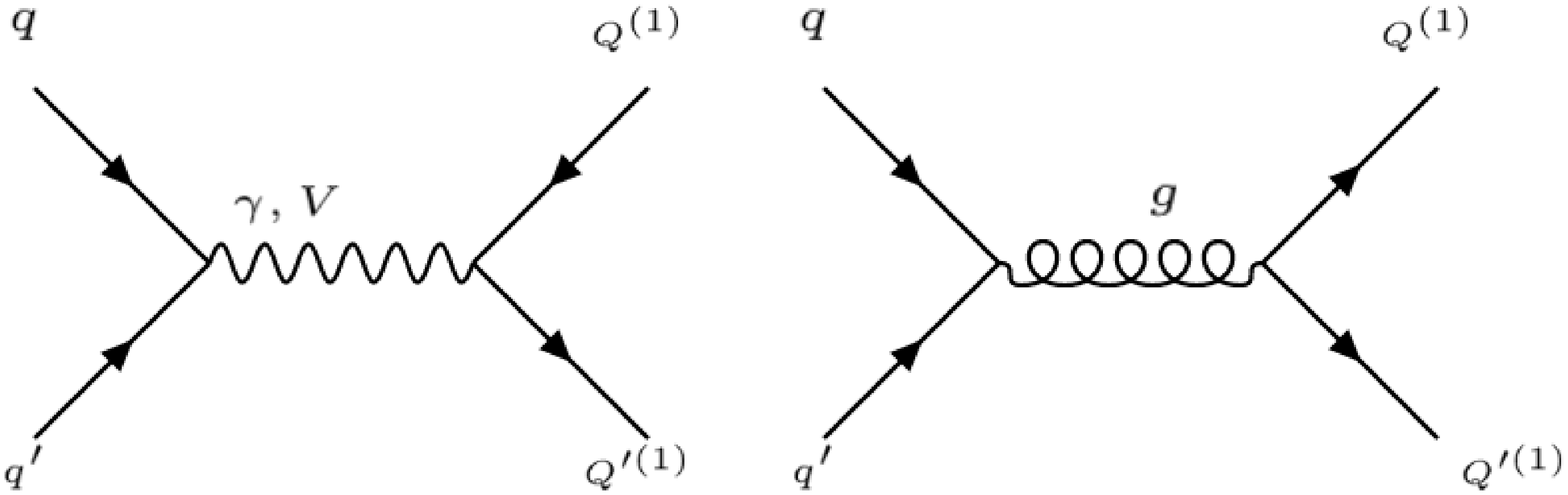}
\includegraphics[width=0.6\textwidth]{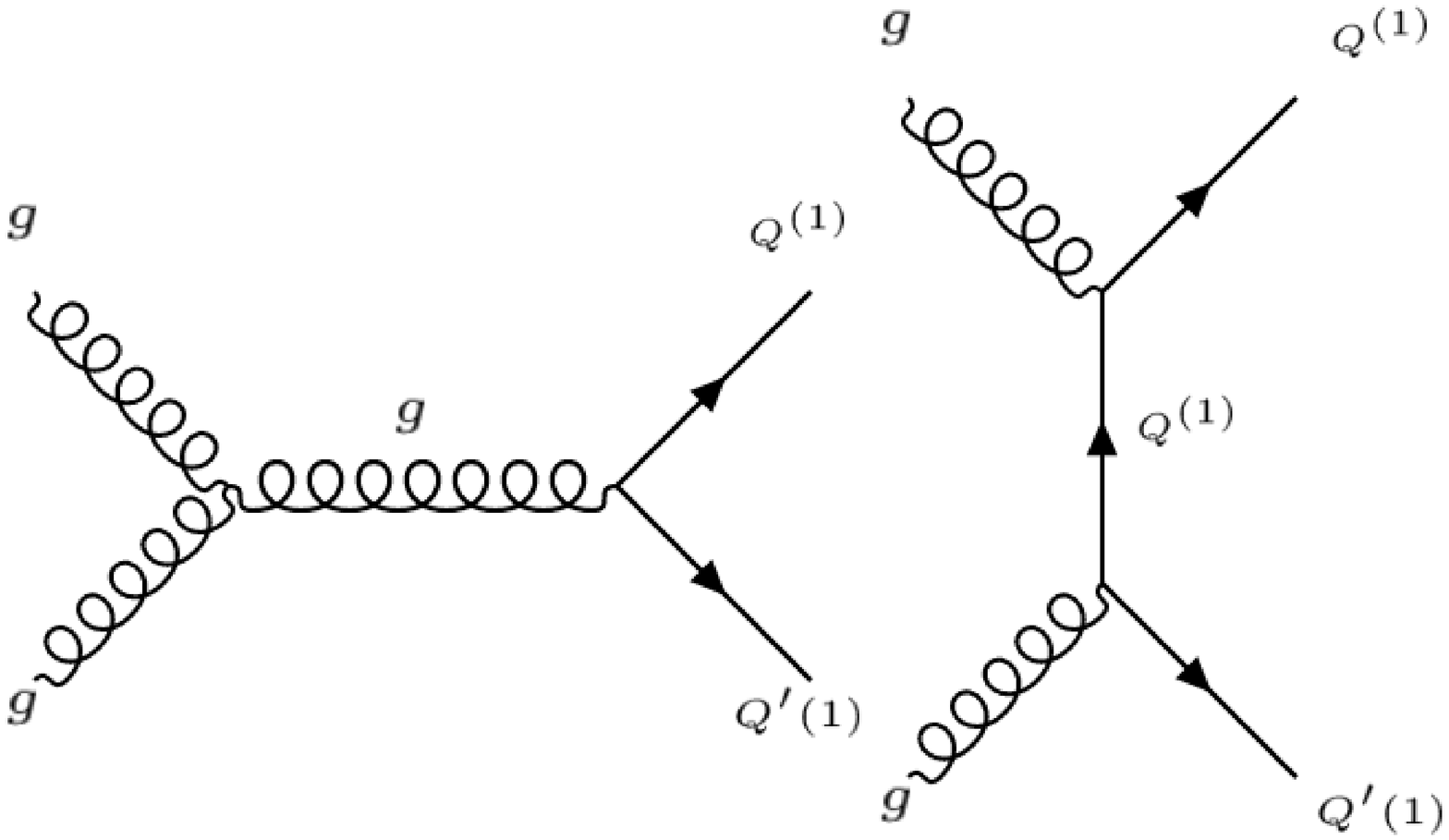}
\includegraphics[width=0.6\textwidth]{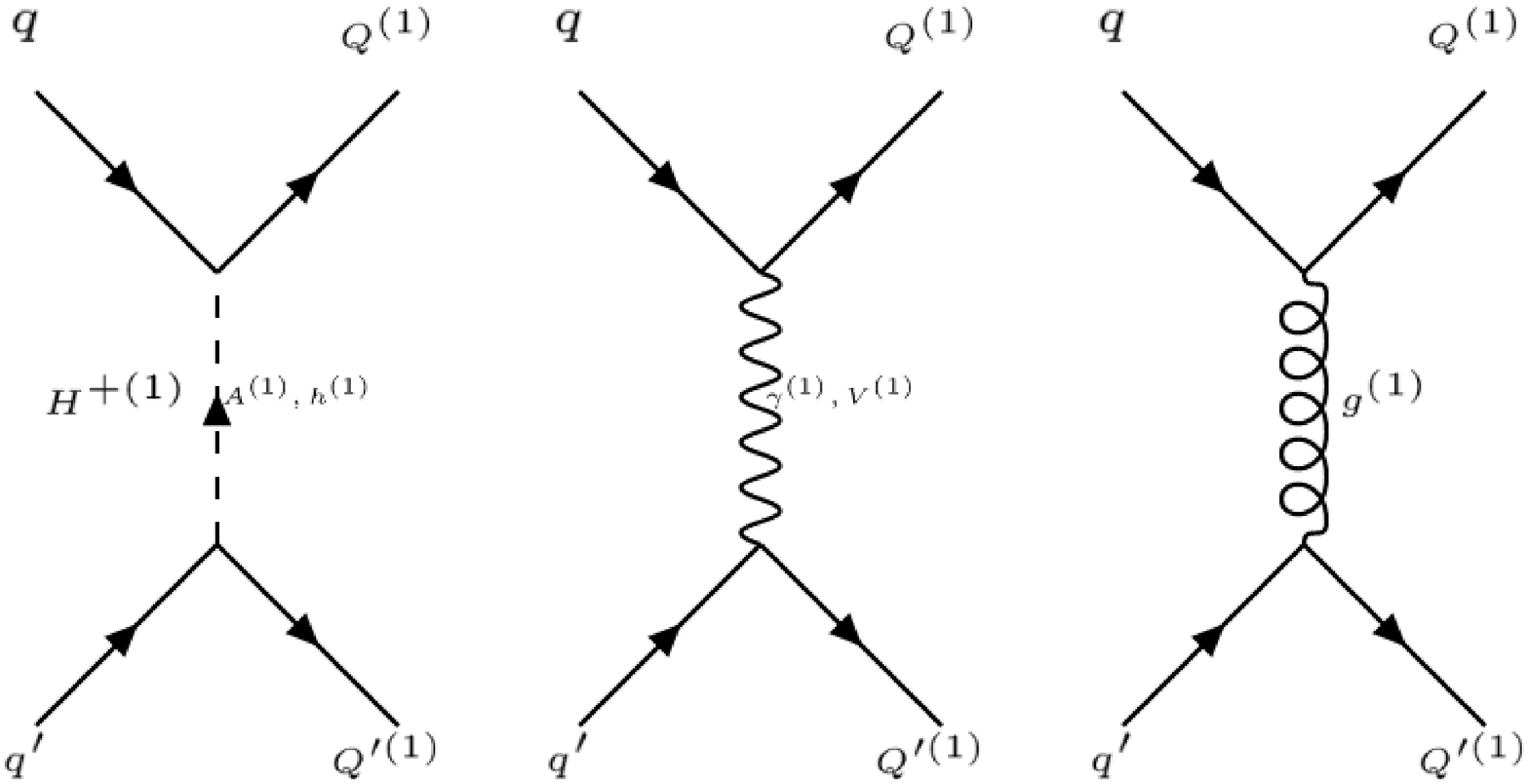}
\caption{Feynman diagrams for pair productions of $n=1$ KK-excitations of quarks. }
\label{fig1}
\end{figure}

\begin{figure}[H]
\centering
\includegraphics[width=0.6\textwidth]{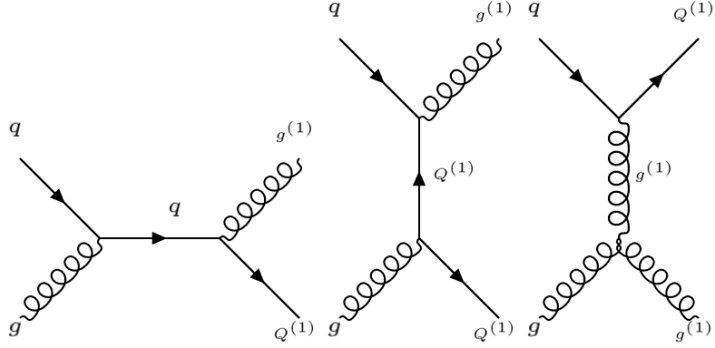}
\caption{Feynman diagrams for associated productions of $n=1$ KK-excitations of quarks and gluons. }
\label{fig2}
\end{figure}

\begin{figure}[h]
\centering
\includegraphics[width=0.6\textwidth]{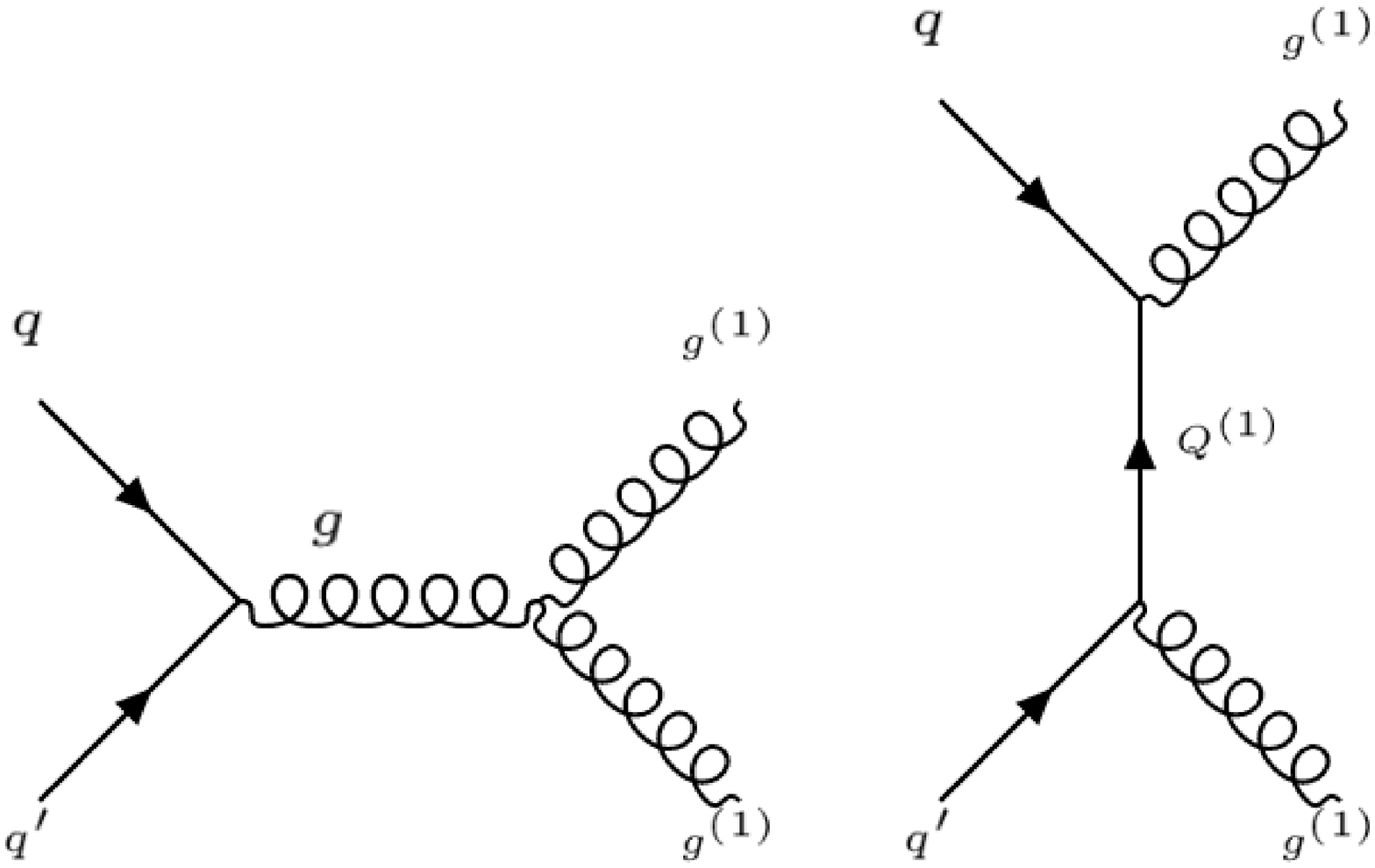}
\includegraphics[width=0.6\textwidth]{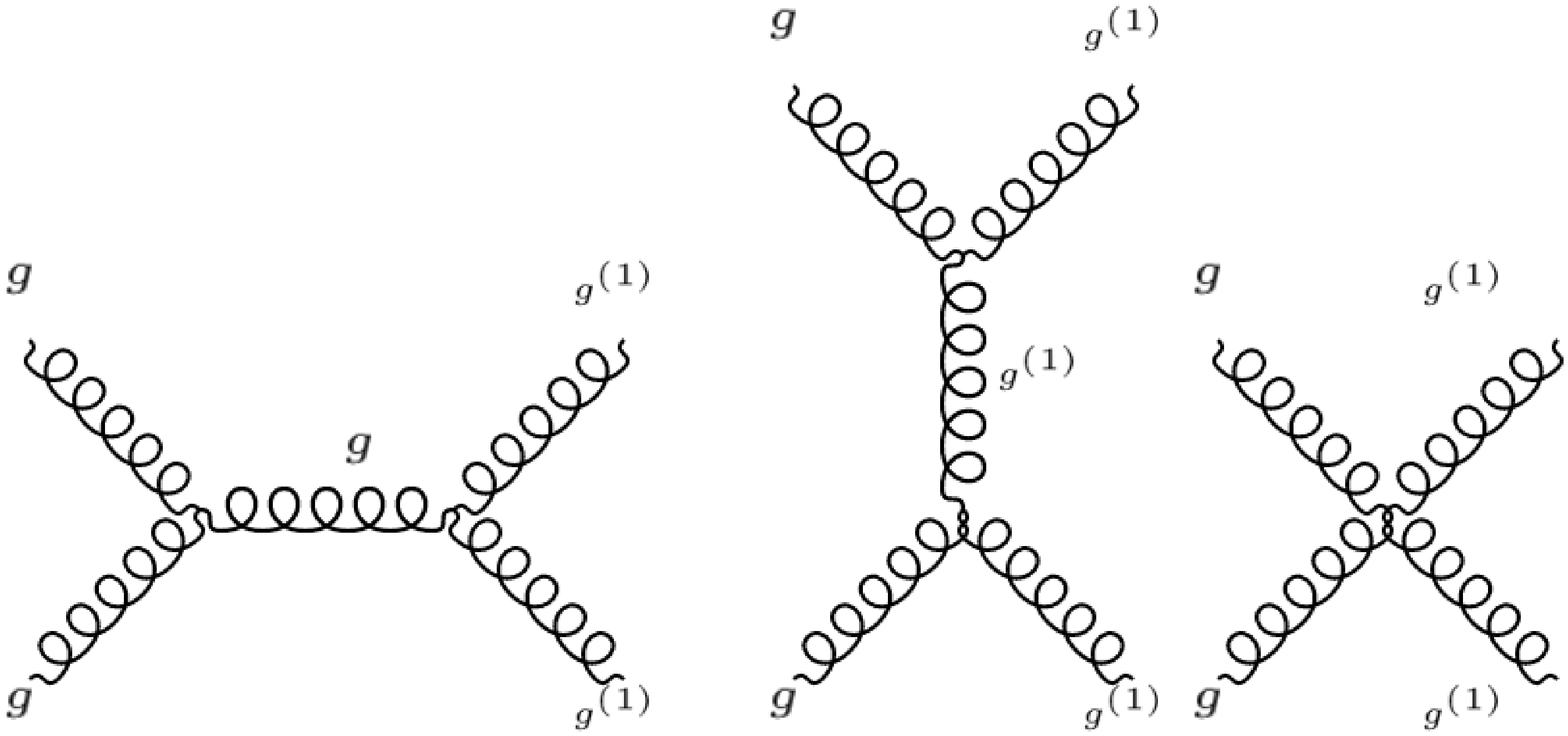}
\caption{Feynman diagrams for pair productions of $n=1$ KK-excitations of gluons. }
\label{fig3}
\end{figure}

\begin{figure}[H]
\centering
\includegraphics[width=0.6\textwidth]{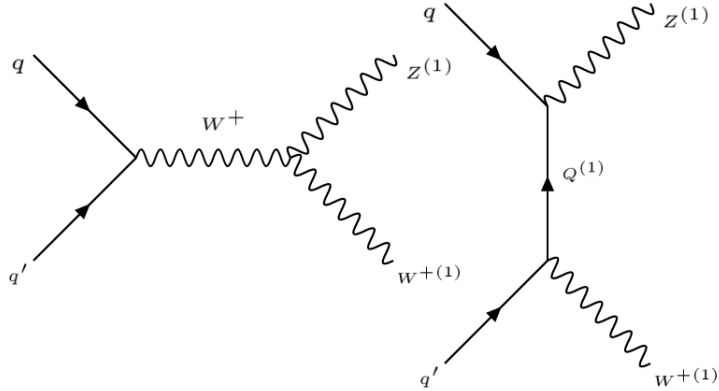}
\caption{Feynman diagrams for pair productions of $n=1$ KK-excitations of EW gauge bosons. }
\label{fig4}
\end{figure}

\begin{figure}[h]
\centering
\includegraphics[width=0.5\textwidth]{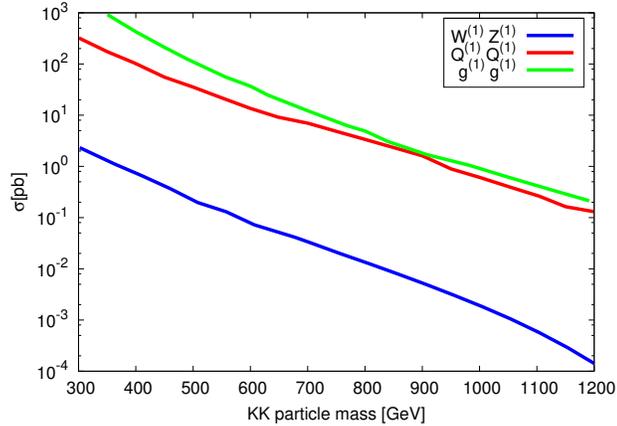}
\caption{Variation of cross-section for the productions of KK particles as a function of their masses at $\sqrt{S} = 13$ TeV. $R^{-1}$ is fixed at 1 TeV.}
\label{fig5}
\end{figure}

In Figs. \ref{fig1}, \ref{fig2} and \ref{fig3}, we present the Feynman diagrams corresponding to the pair and associated productions of KK-quarks and KK-gluons. At the very outset it is to be borne in mind that productions of KK-gluons and KK-quarks are completely determined by $R^{-1}$ which sets the overall mass scale of the KK-excitations along with two BLT coefficients namely $r_{gl}$ and $r_q$ which play the dual role in determining the masses of the KK-excitations and governing the dynamics. However in case of KK-quark pair productions, $r_W$ will also come into play through the $t$-channel exchange diagram (see Fig. \ref{fig1}). The associated EW production of $W^{\pm (1)}$ and $Z^{(1)}$ at the LHC can occur either via a $s$-channel $W^{\pm}$ or $t$-channel $Q^{(1)}$ exchange (see Fig. \ref{fig4}). If mass of $Q^{(1)}$ is large, $t$-channel process will be suppressed compared to the corresponding $s$-channel\footnote{In passing we point out that $Z^{(1)}$ pair production is only driven by a $t$-channel process as in this case $s$-channel diagram is absent. Therefore if $Q^{(1)}$ is very heavy, the cross-section of $Z^{(1)}$ pair production will be very poor as compared to that of $W^{\pm (1)} Z^{(1)}$.}. Like the strong production here also for a fixed value of $R^{-1}$, $r_W$ and $r_q$ determine the EW production cross-section. In Fig. \ref{fig5}, we present cross-sections for both the strong and EW productions for 13 TeV run of the LHC with $R^{-1}$ fixed at 1 TeV. For a $g^{(1)}$ $(Q^{(1)})$ of mass around 400 GeV, the corresponding pair production cross-section is nearly $\sim$ 425 (105) pb. For higher masses $g^{(1)} g^{(1)}$ and $Q^{(1)} Q^{(1)}$, production cross-sections become almost comparable. The NLO corrections to the production cross-sections of strongly interacting KK-particles can be found in \cite{Freitas:2017hov}. However we have not considered these corrections in our analysis. On the other hand, EW cross-section is much smaller as expected.

It is natural to discuss decay patterns of level 1 KK-particles at this point. In the following analysis, we consider $g^{(1)}$ to be the heaviest. Consequently, it decays to $q Q^{(1)}$ with 50$\%$ branching ratio (BR)\footnote{The other half of the time it decays to singlet KK-quarks.}. The decay modes of $Q^{(1)}$, however, depend on the relative values of gauge and fermion BLT coefficients. In Table \ref{tab3}, we show decay patterns of $Q^{(1)}$ with the help of some representative benchmark points (BPs) for several choices of BLT coefficients. To start with we concentrate on the scenario A.  Due to equality of $r_B$ and $r_W$, compositions of  $Z^{(1)}$ and $\gamma^{(1)}$ are exactly similar to the $Z$ and $\gamma$ of the SM. This can be easily seen from eq. \ref{mixingangle}, in which the overlap integral becomes unity in the above limit\footnote{This is due to the normalization of the $y$-dependent functions.}. The mixing angle $(\theta_W ^{(n)})$ between $W_{\mu}^{3(n)}$ and $B_{\mu}^{(n)}$ is same for all KK-levels and  it also equals to the Weinberg angle. We will see very soon that this plays a crucial role in determining the decay patterns of the KK-excitations.
$Q_{u,d}^{(1)}$ can decay to KK-gauge bosons and KK-scalars via two-body decay as $r_{EW} > r_q$. The up-type KK-quark decays dominantly into Cabibbo-favoured mode $q W^{+ (1)}$. However, BRs to $q Z^{(1)}$ and $q \gamma^{(1)}$ are also not negligible. $\gamma^{(1)}$ channel can dominate over the corresponding $Z^{(1)}$ channel for small values of $r_{EW}$. A tiny fraction of $Q_u^{(1)}$ also goes to Cabibbo-suppressed  channel and rarely to $H^{\pm (1)}$. For the down-type KK-quarks, $W^{-(1)}$ mode is dominant and $Z^{(1)}$ mode is sub dominant. The BR in $\gamma^{(1)}$ channel is always smaller than that of $Z^{(1)}$ although it increases with a decrease in $r_{EW}$ for a fixed value of $r_q$. Down-type KK-quark can also decay occasionally into the charged and cp-odd KK-scalars. 
\begin{table}[H]
\begin{center}
\begin{tabular}{|c|c|c|c|c|c||c|c|c|c|c|}
\hline
\hline
BLT &\multicolumn{5}{|c||}{Up type KK-quark, $Q_{u}^{(1)}$} &\multicolumn{5}{c|}{Down type KK-quark, $Q_{d}^{(1)}$}  \\
\cline{2-6} \cline{7-11} 
coefficients &$d W^{+(1)}$ &$u Z^{(1)}$ &$u \gamma^{(1)}$ &$s W^{+(1)}$ &$d H^{+(1)}$ &$u W^{-(1)}$ &$d Z^{(1)}$ &$d \gamma^{(1)}$ &$u H^{-(1)}$ &$d A^{(1)}$       \\
\cline{1-11}
$r_q =$ 4 &0.585 &0.184 &0.174 &0.03 &0.018 &0.621 &0.278 &0.036 &0.041 &0.024 \\
$r_{EW} =$ 19.3 & & & & & & & & & &   \\
\hline
$r_q =$ 4 &0.588 &0.159 &0.213 &0.03 &- &0.682 &0.259 &0.057 &- &- \\
$r_{EW} =$ 4.7 & & & & & & & & & &   \\
\hline
$r_q =$ 0.81 &0.604 &0.194 &0.139 &0.031 &0.023 &0.637 &0.291 &0.035 &0.024 &0.014 \\
$r_{EW} =$ 9.3 & & & & & & & & & &   \\
\hline
$r_q =$ 0.81 &0.608 &0.181 &0.176 &0.031 &- &0.668 &0.282 &0.046 &- &- \\
$r_{EW} =$ 0.98 & & & & & & & & & &   \\
\hline
$r_q =$ 0.35 &0.553 &0.125 &0.291 &0.028 &- &0.689 &0.222 &0.086 &- &-   \\
$r_{EW}  =$ 0.39 & & & & & & & & & &   \\
\hline
$r_q =$ -0.289 &0.621 &0.202 &0.137 &0.032 &- &0.655 &0.302 &0.034 &- &- \\
$r_{EW} =$ 1.7 & & & & & & & & & &   \\ 
\hline
$r_q =$ -0.289 &0.608 &0.181 &0.179 &0.031 &- &0.67 &0.28 &0.047 &- &- \\
$r_{EW} =$ -0.235 & & & & & & & & & &   \\ 
\hline
\hline
\end{tabular}
\end{center}
\caption{Different decay modes of KK-quarks along with their BRs for various choices of BLT coefficients. BLT coefficients are given in TeV$^{-1}$.}
\label{tab3}
\end{table} 
A comparison of first two lines (and consequently third with fourth and sixth with seventh line) would reveal how these BRs depend on $r_q$ and $r_{EW}$. In each of these pairs, $Q_u^{(1)}$ mass remains fixed while KK-gauge boson mass has been increased by decreasing $r_{EW}$ from one BP to the next. $r_{EW}$ not only controls the masses of KK-EW gauge  bosons, it also affects the couplings involved in these decays via overlap integrals which originate at these decay vertices. However, if we compare the first two lines, the effect is mainly kinematical. As $r_{EW}$ changes from 19.3 to 4.7, all the KK-gauge boson masses go up thus squeezing the available phase space for the decays. $Z^{(1)}$ being the heaviest among these three, its mass becomes the closest to that of  $Q_u^{(1)}$ and subsequently $Q_u^{(1)} \rightarrow Z^{(1)} q$ decay width deceases. As a result, decay BR to $\gamma ^{(1)} q$  goes up. While decay rates of $Q_u^{(1)}$ to $Z^{(1)}$ and $\gamma ^{(1)}$ can be quite competitive, in case of $Q_d^{(1)}$ decay to $\gamma^{(1)}$ is always suppressed in comparison to $Z^{(1)}$ as a consequence of electric charge. In a nutshell, decay BRs do not change much over a wide range of BLT coefficients. This is primarily due to the fact that decay widths of $Q^{(1)}_{u,d}$ into different final states are not too sensitive to the couplings through overlap integrals. The variation in BR over the BLT coefficient space which is seen from Table \ref{tab3}, is mainly due to the decay kinematics (mass differences between the decaying and final state KK-particles).
The decay modes of $W^{\pm (1)}$ and $Z^{(1)}$ depend crucially on the hierarchy among different BLT coefficients. When $r_W = r_B$, the EW spectrum is extremely compressed and if KK-leptons are heavier than KK-EW gauge bosons, both $W^{\pm (1)}$ and $Z^{(1)}$ decay via three-body channels. The decay of $Z^{(1)}$ can occur either via $q\bar{q} \gamma^{(1)}/l l \gamma^{(1)}$ or $q q' W^{\pm(1)}/l \nu W^{\pm (1)}$ channel. $W^{\pm (1)}$ being lighter than $Z^{(1)}$, however, always decays into $q q' \gamma^{(1)}/l \nu \gamma^{(1)}$. We illustrate this with the help of few BPs in Table \ref{tab4}.

It can be seen from Table \ref{tab4} that BR of $Z^{(1)}$ into $\gamma^{(1)}$ channel becomes smaller compared to that of $W^{\pm (1)}$ channel as $r_{EW}$ increases. For $r_{EW} \sim 19$, $W^{\pm (1)}$ channel completely overshadows $\gamma^{(1)}$ mode. On the other hand, if we assume $r_W \neq r_B$ (scenario B) and keep the fermion sector below $W^{\pm (1)},Z^{(1)}$, the KK-gauge bosons will decay into fermion-anti fermion pair like their SM counterparts. 
\begin{table}[H]
\begin{center}
\begin{tabular}{|c|c|c|c|c|c|c||c|c|}
\hline
\hline
\multicolumn{3}{|c|}{BLT coefficients} &\multicolumn{4}{c||}{Decay modes of $Z^{(1)}$} &\multicolumn{2}{c|}{Decay modes of $W^{\pm (1)}$}  \\
\cline{1-3} \cline{4-7} \cline{8-9} 
$r_{EW}$ &$r_q$ &$r_l$ &$q q' W^{\pm(1)}$ &$l \nu W^{\pm (1)}$ &$q\bar{q} \gamma^{(1)}$ &$l l \gamma^{(1)}$ &$q q' \gamma^{(1)}$ &$l \nu \gamma^{(1)}$ \\
\cline{1-9}
4.66 &4 &4 &0.093 &0.049 &0.489 &0.366 &0.662 &0.337 \\
\hline
1.165 &1.1 &1.1 &0.029 &0.015 &0.546 &0.408 &0.648 &0.352 \\
\hline
7 &4 &4 &0.434 &0.226 &0.182 &0.136 &0.666 &0.334 \\
\hline
19.3 &3.1 &3.1 &0.653 &0.327 &0.02 &- &0.663 &0.337 \\
\hline
\hline
\end{tabular}
\end{center}
\caption{Different decay modes of KK-EW gauge bosons along with their BRs for various choices of BLT coefficients. BLT coefficients are given in TeV$^{-1}$.}
\label{tab4}
\end{table} 

\section{The Methodology and Results}
\label{method}
We are now ready to present the main results of our analysis. But before doing that we first describe in detail the methods that we have used to constraint the nmUED parameter space in two different scenarios as discussed in section \ref{models}. For all our calculations we restrict ourselves to the production and decay of $n = 1$ KK-modes only and keep $R^{-1}$ fixed at 1 TeV. $R^{-1}$  values beyond 1 TeV will drive the masses of the KK-excitations to such a regime where their production cross-section will be small even at 13 TeV. So we restrain in doing such exercise. In the following we will first discuss how the parameter space in scenario A is constrained from both Run I and Run II LHC data. Then we go on discussing the same for scenario B. We will also present how much parameter space can be explored by looking at the leptonic signal originating mainly from scenario B. The effect of the observed value of DM relic density of the universe and limits coming from the DM direct detection experiments will also be discussed for these scenarios. 

The nmUED model files used in this work are generated using \verb+FeynRules 2.0+ \cite{Alloul:2013bka,Christensen:2008py} running on \verb+Mathematica 9.0+ \cite{mathematica} platform. All the parton level signal events are generated using \verb+MadGraph5_aMC@NLO v.2.4.2+ \cite{Alwall:2014hca} and these events are then passed through \verb+MadGraph5-PYTHIA+ \cite{Sjostrand:2006za} interface for hadronization and showering. We use CTEQ6L \cite{Pumplin:2002vw} parton distribution functions in all our simulations. For the analysis and detector level simulation, all the hadronized events are fed into \verb+CHECKMATE v.1.2.2+ \cite{Drees:2013wra}. 

As already mentioned, nmUED model can provide a DM candidate under the assumption of KK-parity conservation. In this section we will examine the possibility whether the relic density calculated in nmUED model is in agreement with its experimentally measured value from WMAP and PLANCK data. We consider a 3$\sigma$ band of the observed relic density which is consistent with WMAP \cite{Hinshaw} and PLANCK data \cite{Ade:2015xua} : 
\begin{equation}
0.1133 < \Omega h^2 < 0.1265
\end{equation}
In this work we consider $\gamma^{(1)}$ to be a cold dark matter candidate. There are two possible interactions that can contribute to the relic density of DM - annihilations of DM particles into SM particles and co-annihilation between DM candidate and other KK-particles. Now it is well known that the strength of co-annihilation process depends on the mass difference between the DM candidate and the other co-annihilating KK-particle and is important only when there is a nearby KK-excitation \cite{Servant:2002aq,Kong:2005hn}. We have already seen that for a fixed value of $R^{-1}$, the entire KK-spectrum is determined by the BLT coefficients. Therefore the specific choices of BLT coefficients will determine the mass separation of LKP with other KK-particles and that in turn will affect co-annihilation processes.

We have computed the spin-independent LKP-proton scattering cross-section and compare it with the results coming from LUX \cite{Akerib:2016vxi} and XENON1T experiments \cite{Aprile:2017iyp}. The direct scattering between DM and a nucleon can occur either via a $h$-exchange in the $t$-channel or $s$-channel quark exchange. The relative strength of the latter, however, depends on the mass of the quarks involved. DM relic density and DM-proton direct scattering cross-section ($\sigma_{SI}$) have been estimated using \verb+micrOMEGAs+ \cite{Belanger:2013oya}.

\subsection{Constraining scenario A with LHC Run I and Run II data}
\label{modela}
The ATLAS collaboration has looked for final states with isolated leptons ($l$)\footnote{In this work we will always take $l$ to be either $e$ or $\mu$ unless otherwise mentioned.} and jets ($j$) associated with large $\met$ during Run I in the context of squark and gluino searches at the LHC \cite{Aad:2015mia}. We perform the same analysis for production of $n=1$ KK-quarks and KK-gluon in the nmUED framework. $r_f$ and $r_{gl}$ are chosen in such a way so that $m_{gl}^{(1)}$ = 1.4 $m_f^{(1)}$. We allow all the BLT coefficients to vary freely within $-\pi R < r < 20 R$. Assuming  a value of BLT coefficient smaller than $- \pi R$ would  make the corresponding zero-mode tachyonic which can be clearly seen from eq. \ref{normalisation}. The upper bound of $20 R$  arises from the imposition of unitarity  constraints on KK-gauge boson scattering amplitudes \cite{Jha:2016sre}. The hierarchy among BLT coefficients used in this case to scan the parameter space is chosen as $r_{gl} < r_f < r_{EW}$. We consider all possible productions of doublet KK-quark ($Q^{(1)}$) and KK-gluon ($g^{(1)}$) - $Q^{(1)} \bar{Q}^{(1)}$,$Q^{(1)} Q^{(1)}$, $\bar{Q}^{(1)} \bar{Q}^{(1)}$, $g^{(1)} Q^{(1)}$, $g^{(1)} \bar{Q}^{(1)}$ and $g^{(1)} g^{(1)}$. We restrict ourselves to the first two generations of KK-quarks. Decay BRs of KK-quarks and KK-gauge bosons have been discussed in detail in the previous section. However, before delving into the analysis of the LHC signal of this scenario it must be borne in mind that due to small mass separations among the KK-excitations in the EW sector leptons which originate from their decays tend to be soft.

The ATLAS collaboration constructed total nine signal regions (SRs) on the basis of softness or hardness and multiplicity of the leptons and jets in the final state \cite{Aad:2015mia}. These SRs are further binned into transverse momenta of jets ($p_T^{jet}$). A point in the parameter space is said to be excluded if the corresponding number of BSM events ($N_{BSM}$) exceeds the observed upper bound $S^{95}_{obs}$ in any one of the above SRs. We closely follow ref. \cite{Aad:2015mia} for primary selection, isolation and reconstruction of signal objects. Signal $e$s are required to have $|\eta| < 2.47$ and $p_T >$ 7 (10) GeV in the soft (hard) lepton channel. Signal $\mu$ must have a $p_T >$ 6 (10) GeV and $|\eta| < 2.4$ in the soft (hard) lepton channel. Jets are reconstructed using anti-$k_T$ algorithm \cite{Cacciari:2008gp} with radius parameter $R = 0.4$ and must have $p_T > 25$ GeV, $|\eta| < 2.5$. The comparatively soft cut on $p_T$ of lepton is consistent with the compression in the gauge sector. 
  
Next we extend the previous analysis in the case of $\sqrt{S} = 13$ TeV with an integrated luminosity of 36.1 fb$^{-1}$. During Run II, the ATLAS collaboration performed a search for squarks and gluinos with exactly one isolated lepton, jets and large $\met$ in the final state \cite{Aaboud:2017bac}. We use this result to investigate the status of the previous analysis at the upgraded centre of mass (CM) energy. The choice of BLT coefficients is same as before. In this case also we generate all possible combinations of $Q^{(1)}$ and $g^{(1)}$. The decay modes of KK-particles follow exactly the same pattern. In the analysis part, the ATLAS collaboration introduced five SRs. Each SR is characterized by different values of the kinematic cuts. The object selection, isolation and reconstruction are almost same as the previous case and can be found in \cite{Aaboud:2017bac}.

\begin{figure}[h]
\centering
\includegraphics[width=0.5\textwidth]{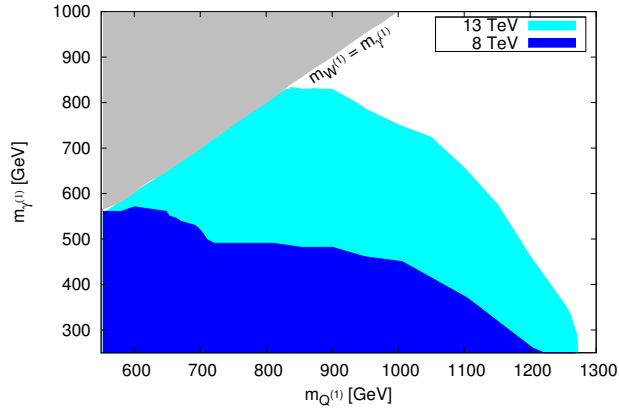}
\caption{The blue and cyan regions represent the excluded parameter space in the $m_{Q^{(1)}} - m_{\gamma^{(1)}}$ plane in case of scenario A obtained by our simulation using the ATLAS data at Run I and Run II respectively. The grey region corresponds to the parameter space which is theoretically excluded as $W^{\pm (1)}/Z^{(1)}$ becomes a LKP there.}
\label{fig6}
\end{figure}

In Fig. \ref{fig6}, we display the result of our scanning in the $m_{Q^{(1)}}$-$m_{\gamma^{(1)}}$ plane for scenario A. The blue region represents the excluded part of the parameter space in this plane by the ATLAS $n l + m j + \met$ search $(n,m \geq 1)$ \cite{Aad:2015mia} at Run I. For $m_{\gamma^{(1)}} \sim 250$ GeV, this search excludes $m_{Q^{(1)}}$ approximately upto 1.2 TeV\footnote{Note that, this puts a bound of 1.7 TeV on $g^{(1)}$ mass which is chosen to be 1.4 times of $m_{Q^{(1)}}$.}. LKP mass cannot be lowered further due to constraint coming from unitarity. For $m_{\gamma^{(1)}} = 600$ GeV and above, there is no bound on $m_{Q^{(1)}}$.

We present the result coming from Run II data in the same plot. The cyan region is excluded from the LHC search in $1l + mj + \met$ channel during Run II \cite{Aad:2015mia}. The bound on $m_{Q^{(1)}}$ is now extended upto nearly 1.26 TeV for $m_{\gamma^{(1)}} = 250$ GeV. This value is not significantly stronger than that coming from the 8 TeV data. Note that, for $R^{-1} =$ 1 TeV, the maximum attainable KK-mass is $\sim$ 1.78 TeV. So, for a $g^{(1)}$ with mass around 1.78 TeV, the mass of $Q^{(1)}$ would be approximately 1.27 TeV (as we set $m_{gl^{(1)}} = 1.4 m_{Q^{(1)}}$). This is the upper bound on $m_{Q^{(1)}}$ in this case and this explains the comparatively small shift along $x$-axis at 13 TeV. However, the total area excluded by Run II data is much larger than that of Run I.

Before closing this subsection, it is worthwhile to mention that in view of KK-parity conservation, $\gamma^{(1)}$ is the lightest KK-particle and a contender to be the candidate for DM. However we have checked explicitly that such a scenario (with $r_B = r_W$) suffers from under abundance of DM at present epoch. This is because along with self-annihilation of LKP, the co-annihilation with $W^{\pm (1)}/Z^{(1)}$ also becomes significant due to the small mass difference between LKP and KK-EW gauge bosons. This in turn makes the relic density much smaller than the experimental values as measured from WMAP \cite{Hinshaw} and PLANCK \cite{Ade:2015xua} data. 

\subsection{Constraining scenario B using LHC data and future projections}
\label{modelb}
Assuming an equality between $r_B$ and $r_W$ would lead to an extremely compressed EW spectra at $n=1$ KK-level. As a result, the decay products of KK-EW gauge bosons $W^{\pm (1)}$ and $Z^{(1)}$ would be extremely soft and thus would evade easily any standard cut-based analysis.  Furthermore, it has been also pointed out that due to the compressed nature of the spectrum, relic abundance of DM is also far lower than the experimentally measured value. So, in such a scenario,  no meaningful limits on their masses could be set either from a collider search or from relic density.  

To see whether any meaningful limit can be put on the masses of the EW KK-excitations from the LHC data or DM relic density, one can move to a scenario (scenario B) where equality between $r_W$ and $r_B$ is not assumed any more. Thus by suitably choosing $r_B$ and $r_W$, considerable mass gap within the EW sector can be realized. The strongly interacting KK-particles are assumed to be much heavier than the EW KK-excitations of $n=1$ KK-level. We set $r_{gl} = -\pi$ TeV$^{-1}$ and $r_q = -2.2$ TeV$^{-1}$ which gives $m_{gl}^{(1)}$ and $m_{Q}^{(1)}$ as 1.78 TeV and 1.68 TeV respectively. KK-leptons are taken to be midway between $W^{\pm (1)}/Z^{(1)}$ and $B^{(1)}$. We set the hierarchy among BLT coefficients in this case as $r_B > r_l > r_W > r_q > r_{gl}$. As a result, both $W^{\pm (1)}$ and $Z^{(1)}$ will decay directly into KK-leptons with 100$\%$ BR. 

A search for the production of charginos and neutralinos is performed by the ATLAS collaboration during Run II in three leptons $(3l)$ channel associated with large $\met$ \cite{Aaboud:2018jiw}. We follow this analysis to put correlated bounds on masses of $W^{\pm (1)} / Z^{(1)}$ and $B^{(1)}$ coming from $3l + \met$ search at the LHC. For this we consider production of $W^{\pm (1)} Z^{(1)}$. Note that while $W^{+ (1)} W^{- (1)}$ production cannot lead to $3l + \met$ final state, $Z^{(1)} Z^{(1)}$ production is highly suppressed by much heavier KK-quarks which appear in $t$-channel diagrams\footnote{See Section \ref{prdctn}.}. There are in total eleven SRs introduced by ATLAS \cite{Aaboud:2018jiw} in the above analysis. The number of observed events in the $3l + \met$ channel in each SR and the corresponding SM background events (see Table 13 and 14 of \cite{Aaboud:2018jiw}) can be used to extract the model independent upper bounds on new physics events in the respective SRs \cite{Choudhury:2017acn}. We use these numbers to find out the excluded region in the $m_{W^{\pm (1)}} - m_{B^{(1)}}$ plane. As before jets are reconstructed using anti-$k_T$ algorithm with radius parameter $R = 0.4$ and must have $p_T > 20$ GeV and $|\eta| < 2.8$. Both $e$ and $\mu$ are required to have $p_T > 10$ GeV  whereas $|\eta| <$ 2.47 (2.5) must be satisfied for $e$ ($\mu$). 
\begin{figure}[thb]
\centering
\subfigure[]{\includegraphics[width=0.45\textwidth]{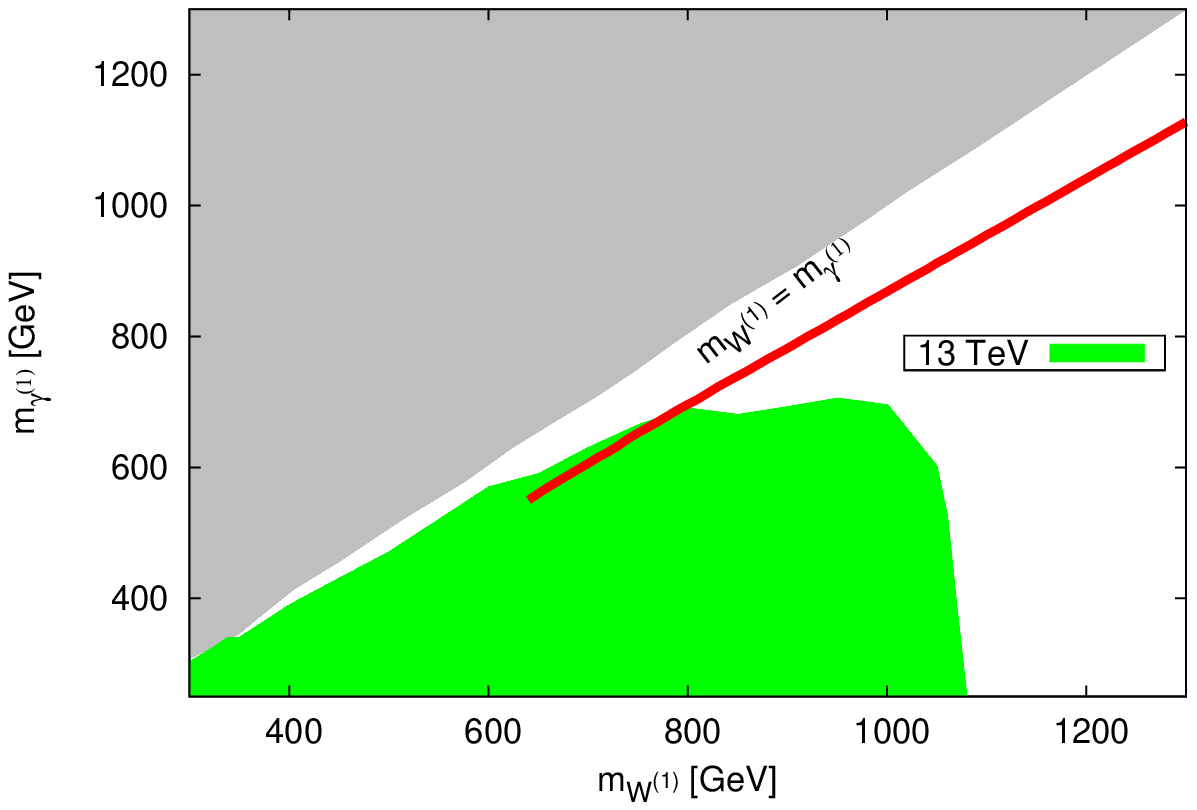}\label{fig7a}}
\subfigure[]{\includegraphics[width=0.45\textwidth]{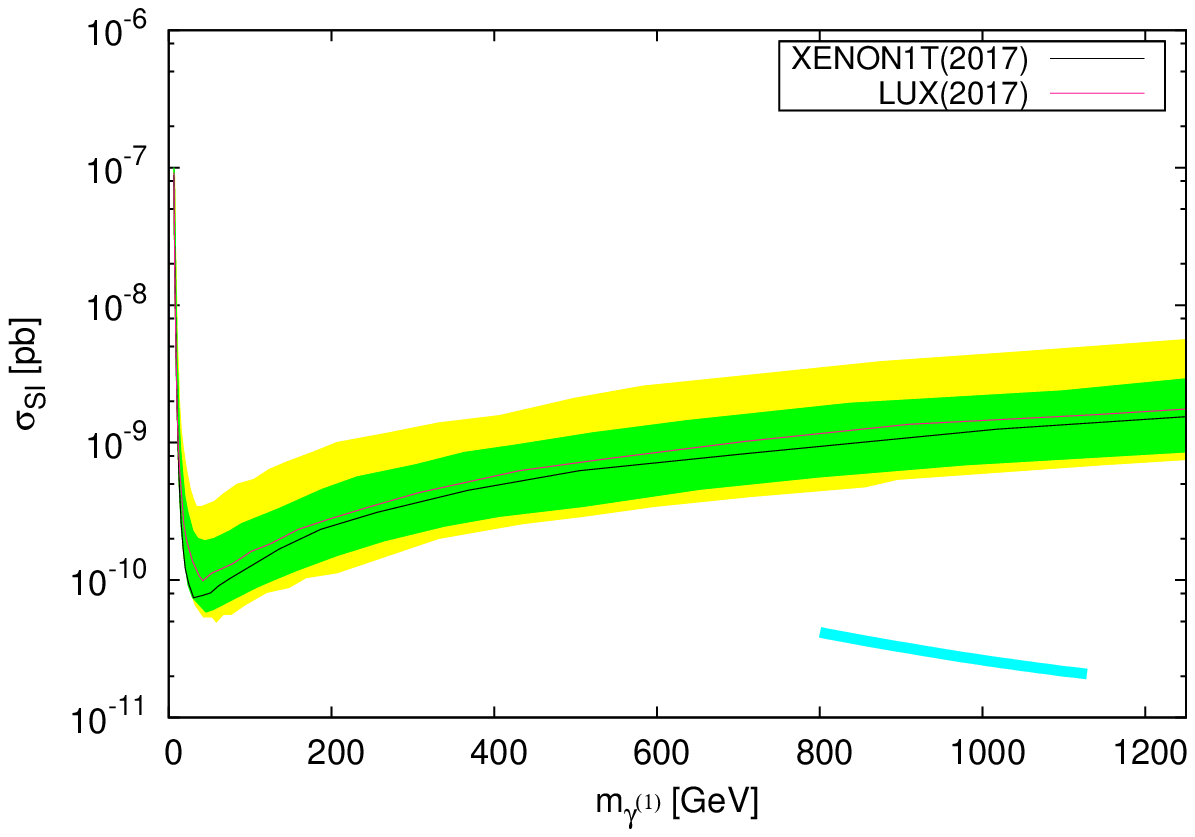}\label{fig7b}}
\caption[]{Left: The green region represents the excluded part of the nmUED parameter space for scenario B in the $m_{W^{\pm(1)}} - m_{\gamma^{(1)}}$ plane coming from $3l + \met$ search during Run II by the ATLAS collaboration. The grey region is theoretically excluded. The red points satisfy WMAP/PLANCK data of DM relic density. 
Right: Direct detection results for scenario B. LUX and XENON1T results are shown as solid pink and black lines respectively. Cyan points satisfy both WMAP/PLANCK data and the LHC constraints. In green and yellow are 1$\sigma$ and 2$\sigma$ sensitivity bands respectively of XENON1T data.}
\label{fig7}
\end{figure}

The allowed parameter space (APS) for scenario B is presented in the $m_{W^{\pm (1)}}$-$m_{\gamma^{(1)}}$ plane of Fig. \ref{fig7a}. The mass of $Z^{(1)}$ which is also controlled by $r_W$ is not shown in the plot\footnote{The mass difference between $W^{\pm (1)}$ and $Z^{(1)}$, however, never exceeds 10 GeV.}. It is evident from Fig. \ref{fig7a} that for $m_{\gamma^{(1)}} = 250$ GeV, $W^{\pm (1)}$ upto 1.1 TeV is ruled out by the LHC trilepton search whereas for $m_{\gamma^{(1)}}$ above 700 GeV all $W^{\pm (1)}$s are allowed. In the same plot, we also show the region (red points) allowed by WMAP/PLANCK data \cite{Hinshaw, Ade:2015xua} assuming the $\gamma^{(1)}$ as the DM candidate. The dominant contribution to the DM relic density comes from LKP pair annihilation into $f \bar{f}$. A tiny fraction goes into $W^{+} W^{-}$ and $Z Z$. Note that, the DM allowed band below $m_{W^{\pm (1)}} = $ 800 GeV is already ruled out by the LHC $3l + \met$ search.
\begin{figure}[h]
\centering
\includegraphics[width=0.5\textwidth]{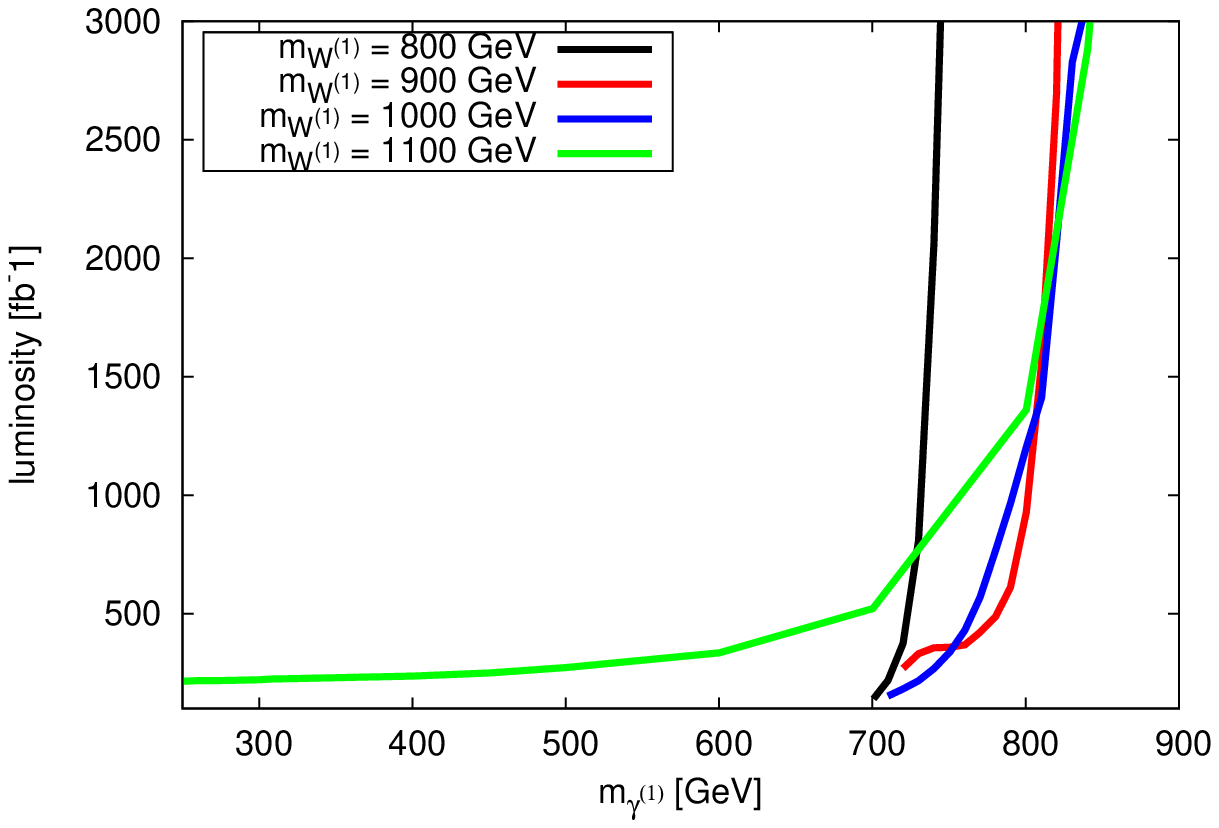}
\caption{Variation of luminosity as a function of $m_{\gamma^{(1)}}$ for $m_{W^{\pm(1)}}$ fixed at four different values - 800, 900, 1000 and
1100 GeV.}
\label{fig8}
\end{figure}

We also examine the possibility of direct detection of DM in this case. The spin-independent scattering of DM with a nucleon can take place either via a $s$-channel KK-quark exchange or $t$-channel $h$ exchange. However, the latter will be dominating over the former due to the $s$-channel suppression arising from heavy KK-quarks. In Fig. \ref{fig7b}, we plot the spin-independent scattering cross-section $\sigma^{SI}$ as a function of $m_{\gamma^{(1)}}$. The exclusion limit coming from latest LUX \cite{Akerib:2016vxi} and XENON1T \cite{Aprile:2017iyp} experiments are shown in pink and black lines respectively where the latter is slightly stronger than the former one. The points consistent with WMAP/PLANCK data and collider limits derived in this section are also shown. As can be seen from Fig. \ref{fig7b} this nmUED scenario is in good agreement with the direct detection data.

Finally, we focus on the prospect of discovering $3l + \met$ signal in scenario B for higher luminosities. We summarize our result in Fig. \ref{fig8}. The four different curves correspond to the variation of required luminosity ($\mathcal L$) as a function of $m_{\gamma^{(1)}}$ for four different masses of $W^{\pm (1)}$. We require the value of signal significance (defined as the ratio $S/\sqrt{B}$ where $S$ and $B$ are the number of signal and background events respectively passing the same sets of cuts) to be at least 5 for discovery and compute the corresponding minimum value of luminosity. The masses of KK-particles are consistent with the bounds coming from $3l + \met$ search (See Fig. \ref{fig7a}). The black line corresponding to $m_{W^{\pm (1)}} =$ 800 GeV is almost vertical i.e. luminosity required to probe this region of APS is higher. This can easily be understood from Fig. \ref{fig4}. The allowed range of $\gamma^{(1)}$ for $m_{W^{\pm (1)}} =$ 800 GeV is 700 - 790 GeV. Therefore despite the comparatively larger cross-section, small mass differences in the EW sector result into low efficiencies and hence higher luminosity is required so that the signal significance becomes at least 5. On the other hand the green line corresponding to $m_{W^{\pm (1)}} =$ 1.1 TeV covers a vast range of LKP masses. Although in this case the cross-section is smaller but comparatively high efficiency makes it possible to get $S/\sqrt{B} =$ 5 for $\mathcal{L} <$ 500 fb$^{-1}$ for a LKP lying in the range 250 - 700 GeV. As the LKP mass increases, efficiency drops and required luminosity for signal to stand over $5\sigma$ fluctuation of the background increases.

\section{Summary and Conclusions}
\label{conclusion}
To summarize,  we have performed an analysis to investigate the  possibility of exploring nmUED model at the LHC experiment. nmUED model is an incarnation of the SM in $4 +1$ space-time dimensions augmented by $S_1 / Z_2$ orbifolding of the extra space-like dimension. The spectrum is hallmarked by the presence of towers of KK-excitations of SM fields. The lightest KK-particle, generally the level one  KK-excitation of $U(1)_Y$ gauge boson, cannot decay as a consequence of imposed $Z_2$ symmetry on the action and fits the bill to be a dark matter candidate. Orbifolding boundary conditions break the momentum conservation along the extra space-like direction. The radiative corrections to the masses which are logarithmically sensitive to the cut-off energy of the theory are localized on the boundaries of the orbifold. However, lack of knowledge about the cut-off forces us to assume rather arbitrary conditions to fix the amount (coefficients of the boundary localized operators) of these corrections. Keeping this in mind one can then also assume the coefficients of these boundary localized operators as free parameters of the theory. and look at the experimental data to fix/constrain  their actual values. This is the philosophy of nmUED. Consequently, masses of the KK-excitations are not exactly equal to the $R^{-1}$ times the KK-number but some other non-integer numbers which carry the imprint of boundary localized terms.

Our main objective in this work has been to look for signatures of nmUED model at the LHC. But before embarking on such road, we have investigated how much parameter space of nmUED has been already explored at the LHC by using the available data. Unfortunately, to best of our knowledge no dedicated search for signals of nmUED models has been done at the LHC. So we have used the existing model independent bounds on new physics coming from other BSM scenarios for our purpose. The particle spectrum in the model of our interest is entirely controlled by compactification radius $R$ and the coefficients of BLTs.  Additionally, these BLT coefficients affect the interactions among KK-excitations via  few overlap integrals and thus play an important role to determine their production cross-sections as well as decay rates.

We have calculated the production cross-sections of the $n=1$ KK-gluons, KK-quarks and KK-EW gauge bosons for $R^{-1} =$ 1 TeV. To understand the relative rates of different final states, we have presented the BRs of different decay modes of these KK-particles emphasizing the subtle role played by BLT coefficients in the decays of KK-particles. Two scenarios of nmUED differing in the hierarchy among the BLT coefficients (and hence among the masses of different KK-excitations) have been considered to study the collider phenomenology. In one scenario, the EW spectrum is chosen to be extremely compressed by setting the $SU(2)_L$ and $U(1)_Y$ BLT coefficients at the same value ($r_B = r_W$). But we allow them to be well separated from the strongly interacting sector comprising of KK-quarks and KK-gluon. We consider the production of KK-gluons and doublet KK-quarks to constraint the corresponding parameters using $nl + mj + \met$ search at both Run I and Run II of the LHC, where lepton(s) coming from decays of KK-EW gauge bosons is essentially soft. We point out that due to the compression in the EW sector, both the self-annihilation and co-annihilation of LKP become significant and hence one is likely to get under abundance of DM in such a scenario. In the second scenario, we assume a less compressed spectrum by choosing a hierarchy like $r_W < r_l < r_B$, among the BLT coefficients. We consider $W^{\pm (1)} Z^{(1)}$ production and use the LHC data of $3l + \met$ search during Run II to put bounds on their masses where leptons originate from the decays of $W^{\pm (1)}$ and $Z^{(1)}$. The existing LHC data can exclude $W^{\pm (1)}$ mass upto 1.1 TeV for $\gamma^{(1)}$ of mass around 250 GeV. A heavier LKP (which means a higher degree of compression among EW KK-excitations) would pull the lower limit on $W^{\pm (1)}$ mass in the opposite direction. We then go on discussing the effects of the measured value of DM relic density of the universe and DM direct detection data on the parameter space of this second scenario. Finally we discuss the discovery prospect of these gauge bosons in the $3l + \met$ channel for higher values of luminosity. Our investigation reveals that $W^{\pm (1)}$ of mass around 1.1 TeV can be explored at 5$\sigma$ when its mass separation with $\gamma^{(1)}$ is at least 800 GeV with integrated luminosity as low as 220 fb$^{-1}$. In our analysis we have not incorporated the NLO corrections to the production of KK-particles. Therefore our result is on the conservative side. 

{\bf Acknowledgements :} Both the authors acknowledge financial support from DST SERB grant. AD and NG thank Sreerup Rayhaudhuri for useful discussions at the initial stage of the work. NG is thankful to Olivier Mattelaer for very helpful discussions on issues with \verb+MadGraph5+.


\end{document}